\documentclass[prl,groupedaddress,reprint,superscriptaddress,floats,nofootinbib,preprintnumbers]{revtex4-1}
\usepackage{graphicx} 
\usepackage{float}
\usepackage{amsmath}
\usepackage{epstopdf}
\usepackage{xspace}
\usepackage{rotating}
\usepackage{ulem}
\usepackage{multirow}
\usepackage{booktabs}
\usepackage[breaklinks=true]{hyperref}
\hypersetup{
  colorlinks=true,
  linkcolor=blue,
  citecolor=blue,
  urlcolor=blue
}

\graphicspath{{./figs/}}
\usepackage{array,tabularx,epsfig,mathrsfs,rotating}
\usepackage{ifthen}
\usepackage{amsfonts}
\usepackage{amssymb}
\usepackage{slashed}

\newcommand{\ETmiss}{\slashed{E}_T}

\renewcommand{\to}{\rightarrow}

\newcommand{\eq}[1]{Eq.~(\ref{#1})}

\newcounter{diagram}

\newcolumntype{C}[1]{>{\centering\let\newline\\\arraybackslash\hspace{0pt}}m{#1}}
\usepackage{xcolor}
\definecolor{Darkgreen}{rgb}{0.,.7,0.2}

\begin{document}

\begin{flushleft} 
IFT-UAM-CSIC-20-124 
\end{flushleft}
\begin{flushright} 
FTUAM-20-17 
\end{flushright}

\title{\boldmath
More light on Higgs flavor at the LHC: \\ 
Higgs couplings to light quarks through $h + \gamma$ production
}
\author{J. A. Aguilar-Saavedra}
\email[]{jaas@ugr.es}
\affiliation{Departamento de F\'{\i}sica Te\'{o}rica y del Cosmos, Universidad de Granada, E-18071 Granada, Spain}
\affiliation{Instituto de Fisica Teorica, IFT-UAM/CSIC,
Cantoblanco, 28049, Madrid, Spain}

\author{J. M. Cano}
\email[]{josem.cano@uam.es}
\affiliation{Instituto de Fisica Teorica, IFT-UAM/CSIC,
Cantoblanco, 28049, Madrid, Spain}
\affiliation{Departamento de Fisica Teorica, Universidad Autonoma de Madrid,
Cantoblanco, 28049, Madrid, Spain
}

\author{J. M. No}
\email[]{josemiguel.no@uam.es}
\affiliation{Instituto de Fisica Teorica, IFT-UAM/CSIC,
Cantoblanco, 28049, Madrid, Spain}
\affiliation{Departamento de Fisica Teorica, Universidad Autonoma de Madrid,
Cantoblanco, 28049, Madrid, Spain
}

\hfill\draft{ }

\begin{abstract}
Higgs production in association with a photon at hadron colliders is a rare process, not yet observed at the LHC. We show that this process is sensitive to significant deviations of Higgs couplings to first and second generation SM quarks (particularly the up-type) from their SM values, and use a multivariate neural network analysis to derive the prospects of the High Luminosity LHC to probe deviations in the up and charm Higgs Yukawa couplings through $h + \gamma$ production.

\end{abstract}

\maketitle


\vspace{-2mm}

\noindent \textbf{I. Introduction.}~Whereas the Yukawa couplings of the 125 GeV Higgs boson to third-generation Standard Model (SM) fermions have been measured rather precisely at the Large Hadron Collider (LHC), the values of the corresponding Higgs boson couplings to light SM fermions are still weakly (or very weakly, for first-generation fermions) constrained.  
In the last few years there has been an important theoretical~\cite{Bodwin:2013gca,Kagan:2014ila,Goertz:2014qia,Perez:2015aoa,Perez:2015lra,Koenig:2015pha,Brivio:2015fxa,Soreq:2016rae,Bishara:2016jga,Bonner:2016sdg,Yu:2016rvv,Cohen:2017rsk,Mao:2019hgg,Coyle:2019hvs,Alasfar:2019pmn} and experimental~\cite{Aad:2015sda,Aaboud:2016rug,LHCb:2016yxg,Aaboud:2017xnb,Aaboud:2018fhh,Sirunyan:2020mds,ATLAS:2020wny} effort to probe the charm quark Yukawa coupling, as well as the rest of the light SM quarks (see e.g.~\cite{Kagan:2014ila,Goertz:2014qia,Soreq:2016rae}).~Some of the proposed methods to probe the Yukawa couplings of the light SM quarks at the LHC are quark-flavor specific (they rely on tagging/identifying a specific flavor in the final state, e.g. a charm quark-jet produced in association with a Higgs boson~\cite{Brivio:2015fxa} or a strange-flavored meson from a rare Higgs decay process~\cite{Kagan:2014ila}), yet others could be sensitive to deviations in any of the Higgs couplings to first and second generation SM quarks. Altogether, there exists a strong interplay among all these different probes, which are key to unravel the details of the mass generation mechanism for the first two generations of matter: 
while the LHC will not be sensitive enough to probe the SM values of the corresponding Higgs Yukawa couplings, it will explore beyond the SM scenarios with significant enhancements in these Yukawa couplings (see~\cite{Porto:2007ed,Giudice:2008uua,Bauer:2015fxa,Bauer:2015kzy,Altmannshofer:2016zrn,Altmannshofer:2017uvs,Egana-Ugrinovic:2019dqu} for some examples).\footnote{Large enhancements of Higgs Yukawa couplings to light quarks can also impact other physical observables, see e.g.~\cite{Bishara:2015cha}.} Our current lack of understanding of the pattern of Higgs Yukawa couplings motivates probing such enhancements to gain insight on the entire Higgs flavor structure, as well as to provide the strongest possible experimental constraints on these couplings (even if still far from the SM predicted values).

\begin{figure}[t]
    \centering
    \includegraphics[width=0.483\textwidth]{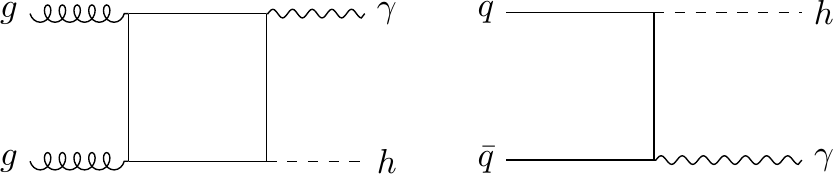}

\vspace{-2mm}
    
    
    
    \caption{Left: Feynman diagram for $g g \to h\gamma$, whose amplitude vanishes due to Furry's theorem. Right: Example tree-level Feynman diagram for $q\bar{q}\to h\gamma$ (with $q = u,d,s,c,b$) in the SM.}
    \label{fig:qqHgamma}
    
    
\end{figure}

In this Letter we explore a complementary  probe of the Higgs couplings to light SM quarks through the production of a Higgs boson in association with a photon at hadron colliders, $p p \to h \gamma$ (see~\cite{Abbasabadi:1997zr,Gabrielli:2007zp,Agrawal:2014tqa,Gabrielli:2016mdd,Arnold:2010dx,Khanpour:2017inb,Dobrescu:2017sue} for other Higgs $+$ photon LHC studies). This is a rare process in the SM, with the leading order (LO) gluon-initiated contribution $g g \to h \gamma$ (see Fig.~\ref{fig:qqHgamma}--left) vanishing due to Furry's theorem~\cite{Furry:1939,PeskinSchroeder}. The largest contributions to the inclusive $h \gamma$ production at the LHC include extra objects with high transverse momentum in the final state~\cite{Gabrielli:2016mdd}. In the absence of such extra final-state particles besides the Higgs boson and photon, the contribution to Higgs $+$ photon production at the LHC from bottom-antibottom ($b \bar{b}$) and charm-anticharm ($c \bar{c}$) initial states (see Fig.~\ref{fig:qqHgamma}--right) becomes important, making this process sensitive to the respective Higgs Yukawa couplings $y_b$ and $y_c$. In addition, the presence of a large deviation from its SM value in the Yukawa couplings of the quarks $q = s,u,d$ (strange, up and down) would greatly enhance the corresponding $q\bar{q}$-initiated contribution from Fig.~\ref{fig:qqHgamma}--right.
These contributions are at the same time proportional to 
the square of the quark electric charge $Q_q$, which suppresses the cross section for down-type quark-initiated $q\bar{q}\to h\gamma$ processes relative to up-type quark-initiated by a factor $(Q_u/Q_d)^2 = 4$.
We thereby 
study the sensitivity of this process to the value of the Yukawa coupling $y_{q}$ for $q = u,c$ at the High-Luminosity (HL) LHC, focusing on (in our view) the most promising Higgs decay channel for this purpose, $h \to W W^* \to \ell\nu\ell\nu$ ($\ell$ being electrons/muons).


\vspace{2mm}

\noindent \textbf{II. $h + \gamma$ production at LHC.}~As outlined in the introduction, the dominant $q\bar{q}$-initiated contributions to the exclusive production of a 125 GeV Higgs boson in association with a photon at hadron colliders (see Fig.~\ref{fig:qqHgamma}--right) are proportional to the square of the corresponding light quark Yukawa coupling $y_{q}^2$, evaluated at the scale of the Higgs mass $m_h$. 
The running masses for the bottom, charm and up quarks, evaluated at the scale $m_h = 125$ GeV, are given in the tadpole-free pure $\overline{MS}$ scheme by $m_b(m_h) = 2.777~\textrm{GeV}$, $m_c(m_h) = 0.605~\textrm{GeV}$, $m_u(m_h) = 0.0013~\textrm{GeV}$~\cite{Martin_2019}, with the SM values of the Yukawa couplings at this scale given by $y^{\mathrm{SM}}_q (m_h) = \sqrt{2}\,m_q(m_h)/v$ and $v$ the electroweak (EW) scale.
We then parametrize the departure of the Higgs Yukawa couplings to light quarks from their SM values as $\kappa_q = y_q(m_h)/y^{\mathrm{SM}}_q (m_h)$. 

\vspace{1mm}

The respective $\sqrt{s} = 14$ TeV center of mass (c.o.m.) LHC cross sections at LO for $b\bar{b}\to h\gamma$, $c\bar{c}\to h\gamma$ and $u\bar{u}\to h\gamma$ evaluated with \textsc{MadGraph 5}~\cite{Alwall:2014hca},
for a photon with transverse momentum $p_T^\gamma > 20~\textrm{GeV}$ and pseudorapidity $|\eta^\gamma|<2.5$, using the \textsc{NNPDF31$\_$nnlo$\_$as$\_$0118$\_$luxqed}~\cite{Bertone:2017bme} parton distribution functions (PDF) set, are
\begin{equation}
\begin{gathered}
\label{XS_bcu}
\sigma_{b\bar{b}} = \kappa_b^2 \times 0.397~\textrm{fb}  \quad,\quad  \sigma_{c\bar{c}} = \kappa_c^2 \times 0.160~\textrm{fb}, \\ 
\sigma_{u\bar{u}} = \kappa_u^2 \times 5.16 \times 10^{-3}~\textrm{ab} \, .
\end{gathered} 
\end{equation}
For the SM, the $c\bar{c}$ contribution is found to be smaller but comparable to $\sigma_{b\bar{b}}$ (despite the large hierarchy between Yukawa couplings), owing to the relative $(Q_c/Q_b)^2=4$ factor and larger PDF of the charm quark w.r.t the bottom. 
At the same time, while $\sigma_{u\bar{u}}$ in the SM is negligible, an enhancement of the up-quark Yukawa making it comparable to the SM charm Yukawa $y_u(m_h) \sim y^{\mathrm{SM}}_c (m_h)$ (corresponding to $\kappa_u \sim 500$) would raise the $u\bar{u}$-initiated $h\, \gamma$ cross section to $\sim 1.3$ fb\footnote{This is a factor $\sim 10$ larger than the SM value for $\sigma_{c\bar{c}}$ from~\eqref{XS_bcu} due to the much larger PDF for the up-quark inside the proton.}. This might allow for a test of first {\sl vs} second generation Yukawa universality in the up quark sector at HL-LHC with $3$ ab$^{-1}$ of integrated luminosity via this process.
We also note that subdominant contributions to the $q \bar{q} \to h\, \gamma$ exclusive production, such as 
$q \bar{q} \to \gamma^* / Z^* \to h\, \gamma$, quickly become negligible for sizable light Yukawa enhancements, e.g.~for $\kappa_c \sim 3$ their size is $\sim 5\%$ of the $\sigma_{b\bar{b}} + \sigma_{c\bar{c}}$ cross section sum.

\vspace{1mm}

Before presenting our analysis in the next section, let us discuss briefly the production of a Higgs boson and a photon at the LHC in an inclusive manner, allowing for extra high-$p_T$ objects to be produced in the process. The dominant contributions to the inclusive $h + \gamma$ production are~\cite{Gabrielli:2016mdd,Arnold:2010dx} vector boson fusion (VBF, $h \gamma j j$) and associated production with a $W$ or $Z$ boson (AP, $h\gamma V$). Slightly smaller than the latter but also important are the production together with a high-$p_T$ jet ($h \gamma j$) and production in association with a top quark pair ($t \bar{t} h \gamma$). 
Cross sections for these processes are in the $\mathcal{O}(1 - 10)$ fb ballpark, and they do not depend on $\kappa_q$ (except for small contributions to $h \gamma j$ and $h \gamma j j$, only important for large $\kappa_c$ values). Thus, to gain sensitivity to the Higgs Yukawa couplings to light quarks, these processes need to be efficiently suppressed 
in favor of the $b \bar{b}$ and $c \bar{c}$-initiated ones. Fortunately, this may be easily achieved by vetoing extra hard activity in the $h\,\gamma$ event selection and exploiting the different kinematics of the Higgs boson and photon among these processes, as we will discuss below.



\vspace{2mm}

\noindent \textbf{III. Sensitivity via $h \to W W^* \to \ell \nu \ell \nu$.}~In the remainder of this work we focus on the $h \to W W^* \to \ell^+ \nu \ell^- \bar{\nu}$ decay of the Higgs boson as the most sensitive channel for our purposes. 
Other Higgs decay choices like $h \to b \bar{b}$ and $h \to \tau^+\tau^-$ face very large SM backgrounds, 
or suffer from very small decay branching fractions, as is the case of $h \to \gamma\gamma$ and $h \to Z Z^* \to 4\ell$. 


\vspace{1mm}

To search for the $h\,\gamma$ signature via the decay $h \to W W^* \to \ell^+ \nu \ell^- \bar{\nu}$ at the LHC with $\sqrt{s} = 14$ TeV c.o.m.~energy, we select events with exactly two oppositely charged leptons (electrons or muons) and a photon with pseudorapidities $|\eta^{\ell , \gamma}| < 4$. The transverse momentum of the photon is required to satisfy $p^{\gamma}_T > 25$ GeV, and the transverse momenta of the leading ($\ell_1$) and subleading ($\ell_2$) lepton need to satisfy $p^{\ell_1}_T > 18$ GeV, $p^{\ell_2}_T > 15$ GeV or $p^{\ell_1}_T > 23$ GeV, $p^{\ell_2}_T > 9$ GeV, following Run-2 ATLAS di-lepton triggers~\cite{ATL-DAQ-PUB-2018-002}. 
Di-lepton trigger thresholds are in fact expected to lower for HL-LHC~\cite{TriggerHL-LHC}, and a di-lepton $+$ photon trigger with lower thresholds could also be implemented.
We also require the missing transverse energy in the event to be 
$\slashed{E}_{T} > 35$ GeV. In order to suppress events with extra high-$p_T$ activity, we veto events having a jet with $p_T > 50$ GeV or 
having two jets with $p_T > 20$ GeV and a pseudorapidity gap $\Delta \eta^{j_1 j_2} > 3$.

The dominant SM backgrounds are the irreducible processes $p p \to \ell^+\nu\ell^-\bar{\nu}\gamma$ and $p p \to Z \gamma$, $Z \to \tau^+ \tau^-$ with both $\tau$-leptons decaying leptonically, together with the reducible background $p p \to t\bar{t} \gamma$ (with $t \to b \ell^+ \nu$, $\bar{t} \to \bar{b} \ell^- \bar{\nu}$). The latter can be further suppressed by imposing a $b$-tagged jet veto on the selected events. We note that the 
$Z\, +$ jets and $Z (\to \ell\ell) \gamma$ SM backgrounds have a very large cross section 
(see e.g.~\cite{Aaboud:2017hbk,Sirunyan:2018cpw,Aad:2019gpq}). However, the above 
selection, in particular the $\slashed{E}_{T}$ cut, combined with a $Z$-mass window veto on the invariant mass of the two leptons $ \left|m_Z - m_{\ell\ell}\right| > 30$ GeV greatly suppresses these processes. Selecting the two leptons in the event to be of opposite flavor (OF) would provide an additional suppression for these backgrounds.   
In any case, we retain both OF and SF (same flavor) lepton events\footnote{Considering only OF events results in a $\sim \sqrt{2}$ reduction in our signal sensitivity. Yet, an experimental analysis splitting the events into OF and SF categories would recover part of this sensitivity. We also note that the SF signal events contain a minor contribution from $h \to Z Z^* \to \nu\bar{\nu} \,\ell^+ \ell^-$.}, and disregard $Z\, +$ jets and $Z(\to \ell\ell)\gamma$ backgrounds altogether.

We generate our signal and SM background event samples (both at LO) in {\sc MadGraph 5}~\cite{Alwall:2014hca} with subsequent parton showering and hadronization with {\sc Pythia 8}~\cite{Sjostrand:2014zea} and detector simulation via {\sc Delphes v3.4.2}~\cite{deFavereau:2013fsa}, using the anti-$k_T$ algorithm~\cite{Cacciari:2008gp} with $R =0.4$ for jet reconstruction with 
{\sc FastJet}~\cite{Cacciari:2011ma} and the {\sc Delphes} detector card designed for HL-LHC studies. We do not include pile-up in our simulation for simplicity: in the experimental measurements, it has been shown that the pile-up contamination can be very efficiently removed by using pile-up subtraction algorithms such as {\scshape Puppi}~\cite{Bertolini:2014bba}, {\scshape Softkiller}~\cite{Cacciari:2014gra} or constituent level subtraction~\cite{Berta:2014eza}.

After event selection, the SM background cross sections are $5.08$ fb for $p p \to \ell^+\nu\ell^-\bar{\nu}\gamma$, $3.86$ fb for $Z \gamma$, $Z \to \tau^+\tau^-$ and $1.07$ fb for $t \bar{t} \gamma$, where the latter includes the effect of the various vetoes in the selection. Assuming SM branching fractions for the Higgs boson (we discuss variants of this assumption in the next section), the signal cross section after event selection is $27.6$ ab for $\kappa_b = \kappa_u = 1$, $\kappa_c = 10$, and 
$41.2$ ab
for $\kappa_b = \kappa_c = 1$, $\kappa_u = 2000$.
In the following, we consider independently the possible enhancement of the charm and up-quark Yukawa couplings w.r.t.~their SM values, performing two separate sensitivity studies. 

\begin{figure}[t]
\begin{centering}
\includegraphics[width=0.483\textwidth]{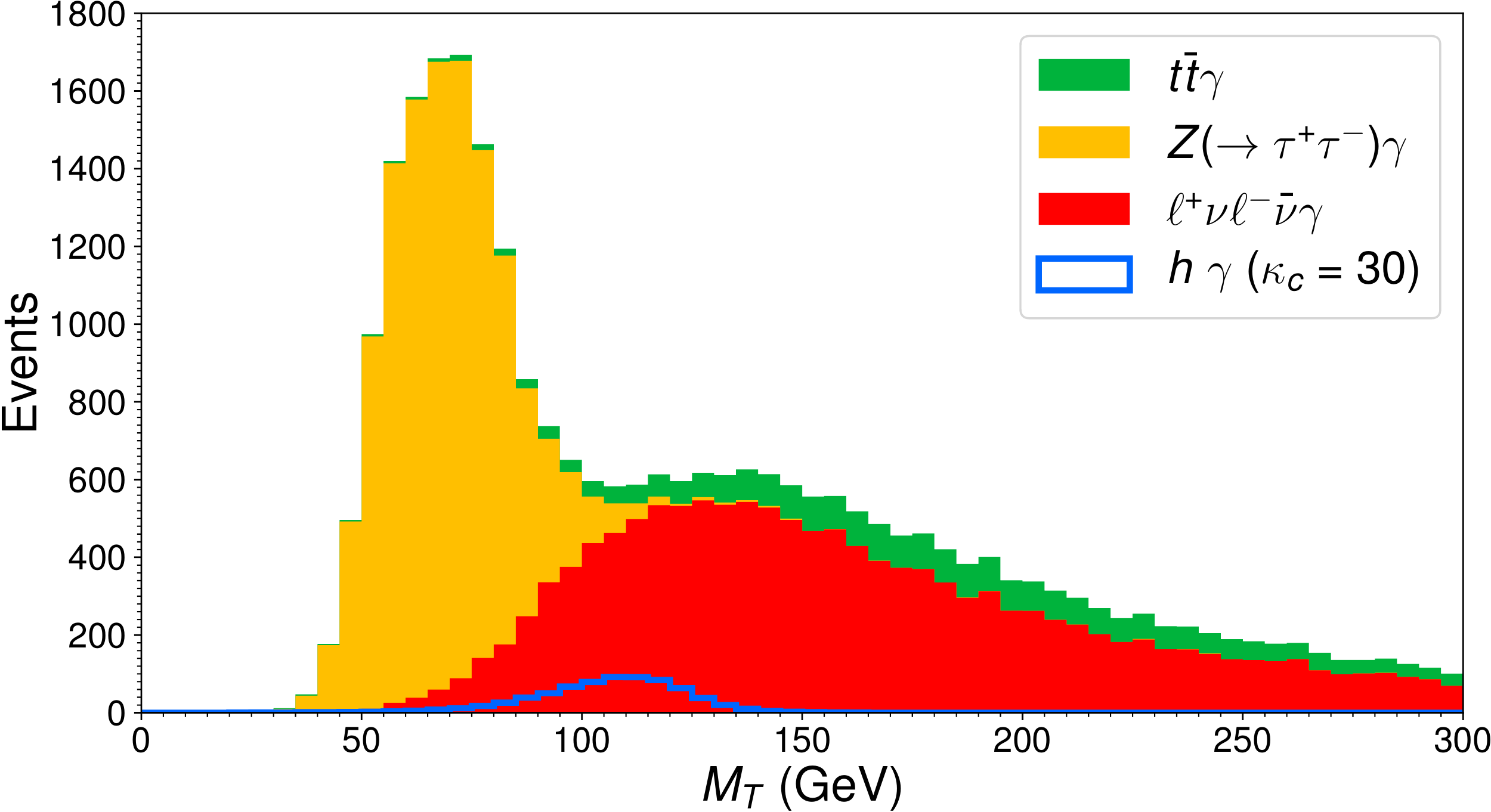}

\vspace{0.5mm}
\includegraphics[width=0.483\textwidth]{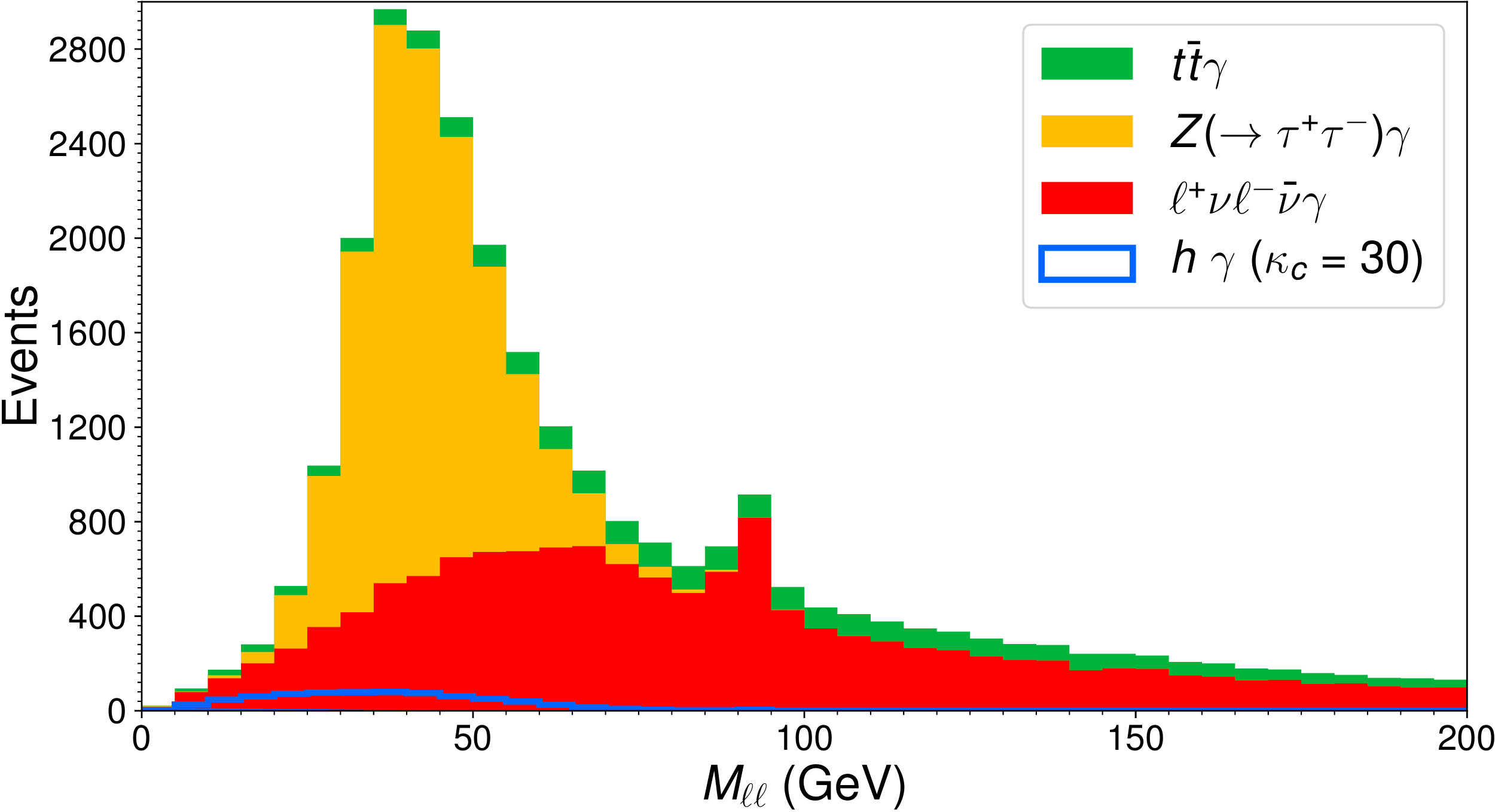}

\vspace{0.5mm}
\includegraphics[width=0.483\textwidth]{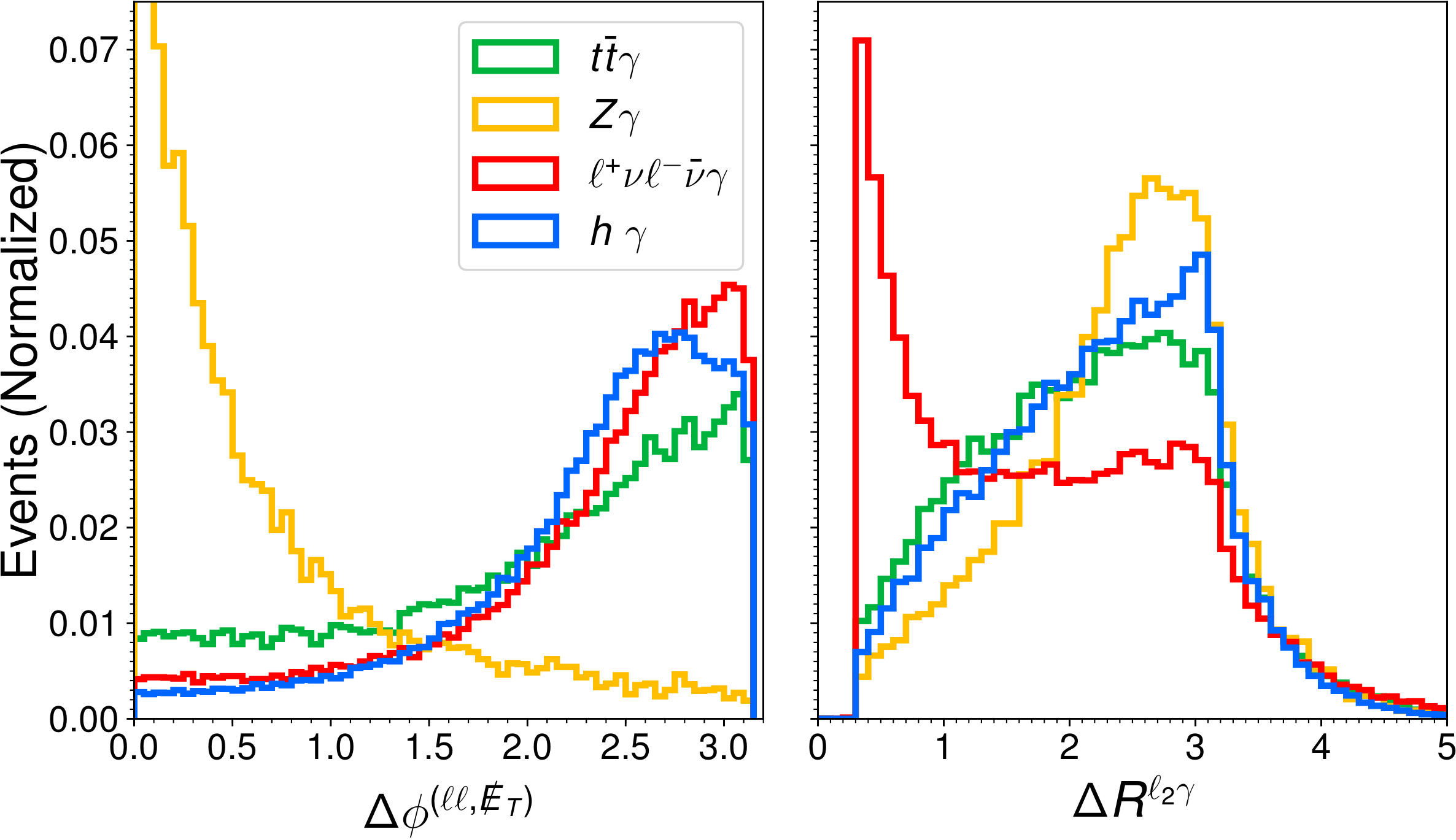}

\vspace{-2mm}

\caption{Top: $M_T$ distribution of events for the dominant SM backgrounds $\ell^+\nu\ell^-\bar{\nu}\gamma$ (red), $t \bar{t} \gamma$ (green), and $Z (\to \tau^+ \tau^-) \gamma$ (yellow), all stacked, at the HL-LHC ($\sqrt{s} = 14$ TeV, 3 ab$^{-1}$). In blue the corresponding $M_T$ distribution for the $h \gamma$ signal with $\kappa_b = \kappa_u = 1$, $\kappa_c = 30$. Middle: same as above, but for $M_{\ell\ell}$ variable. Bottom: Normalized $\Delta \phi^{(\ell\ell,\ETmiss)}$ and $\Delta R^{\ell_2 \gamma}$ distributions for signal and SM backgrounds.}
\label{fig2:MT}
\par\end{centering}


\end{figure}

The rich event kinematics allows for an efficient signal discrimination following the 
initial event selection discussed 
above.~
An important role is played by the transverse mass $M_T$ reconstructed out of the di-lepton system $+$ missing energy:
\begin{equation}
    M_T^2 = \Big( \sqrt{ M_{\ell\ell}^2 + |\vec{p}_T^{\,\,\ell\ell}|^2} + \ETmiss\Big)^2 -
        \left| \vec{p}_T^{\,\,\ell\ell} + \vec{\slashed{E}}_T \right|^2\,,
    \label{eq:MT}
\end{equation}
with $\vec{p}_T^{\,\,\ell\ell}$ the vector sum of the lepton transverse momenta, 
$M_{\ell\ell}$ the invariant mass of the di-lepton system and $\vec{\slashed{E}}_T$ the missing transverse momentum of the event. 
Other key variables are the di-lepton invariant mass $M_{\ell\ell}$ itself, the 
transverse angular separation $\Delta \phi^{(\ell\ell,\ETmiss)}$ between di-lepton momentum $\vec{p}_T^{\,\,\ell\ell}$ and missing momentum $\vec{\slashed{E}}_T$, or the distance $\Delta R \equiv \sqrt{\Delta \phi^2 + \Delta \eta^2} $ between each lepton and the photon $\Delta R^{\ell_1 \gamma}$, $ \Delta R^{\ell_2 \gamma}$. In Fig.~\ref{fig2:MT} we show the $M_T$ (top) and $M_{\ell\ell}$ (middle) distributions for the signal 
(with $\kappa_b = \kappa_u = 1$, $\kappa_c = 30$) and the dominant SM backgrounds at the HL-LHC. We also show in 
Fig.~\ref{fig2:MT} (bottom) the normalized $\Delta \phi^{(\ell\ell,\ETmiss)}$ and $\Delta R^{\ell_2 \gamma}$ distributions for the signal and SM backgrounds. Performing a cut-and-count signal selection
$M_{T} \in [80, 150]$ GeV, $M_{\ell\ell} \in [5, 55]$ GeV, $\Delta R^{\ell_1 \gamma} > 1$, $\Delta R^{\ell_2 \gamma} > 0.8$ and $\Delta \phi^{(\ell\ell,\ETmiss)} > 2$ allows to extract a HL-LHC projected sensitivity 
$|\kappa_c| < 13.9$ at 95\% confidence level (C.L.), using a simple $S/\sqrt{B} \simeq 2$ estimate (with $S$ and $B$ the number of signal and background events) and assuming Higgs boson SM branching fractions.



Given the variety of relevant event kinematic variables and the significant correlations among several of them, it is possible to enhance the signal sensitivity w.r.t.~the above ``squared'' cut-and-count analysis by accessing the full kinematic information of the events. To this end, we adopt here a multivariate approach,  
and use the following set of kinematic variables (which contains all the relevant kinematic information of each event)
\begin{align}
& M_T \,, M_{\ell \ell} \,, M_{\ell \ell \gamma} \,,
p_T^{\,\,\ell_1} \,, p_T^{\,\,\ell_2} \,,
p_T^{\,\,\gamma} \,, \ETmiss \,, \notag \\
& \Delta \phi^{\ell \ell} \,, \Delta \phi^{\ell_1 \gamma} \,,
\Delta \phi^{\ell_2 \gamma} \,, \Delta \phi^{(\ell\ell,\ETmiss)} \,,
 \eta^{\ell_1} \,,  \eta^{\ell_2} \,,  \eta^{\gamma} \,, 
 \label{NN_variables}
\end{align}
to train a neural network (NN) to discriminate the $h\,\gamma$ signal from the various SM backgrounds. The NN architecture uses two hidden layers of 128 and 64 nodes, with Rectified Linear Unit (ReLU) activation for the hidden layers and a sigmoid function for the output layer. The NN is optimized using as loss function the binary cross-entropy, using the Adam optimizer~\cite{kingma2014adam} (other generalized loss functions such as the one proposed in~\cite{Murphy:2019utt} do not give an appreciable improvement). Since the experimental dataset is unbalanced, that is, the SM background overwhelms the signal, it is useful to train the NN using more SM background than signal events, so that the NN learns optimally to identify (and reject) the former. Specifically, we use $1.5 \times 10^4$ events for the $\ell^+\nu\ell^-\bar{\nu}\gamma$ background, $10^4$ events for the $t \bar t \gamma$  background and $5000$ events for the $Z\gamma$ ($Z \to \tau^+ \tau^-$) background (a total of $3\times 10^4$ SM background events) in the NN training, together with $1.5 \times 10^4$ events of $h\,\gamma$ signal. The validation set contains the same number of events from each class.

\begin{figure}[h]
\begin{centering}
\includegraphics[width=0.483\textwidth]{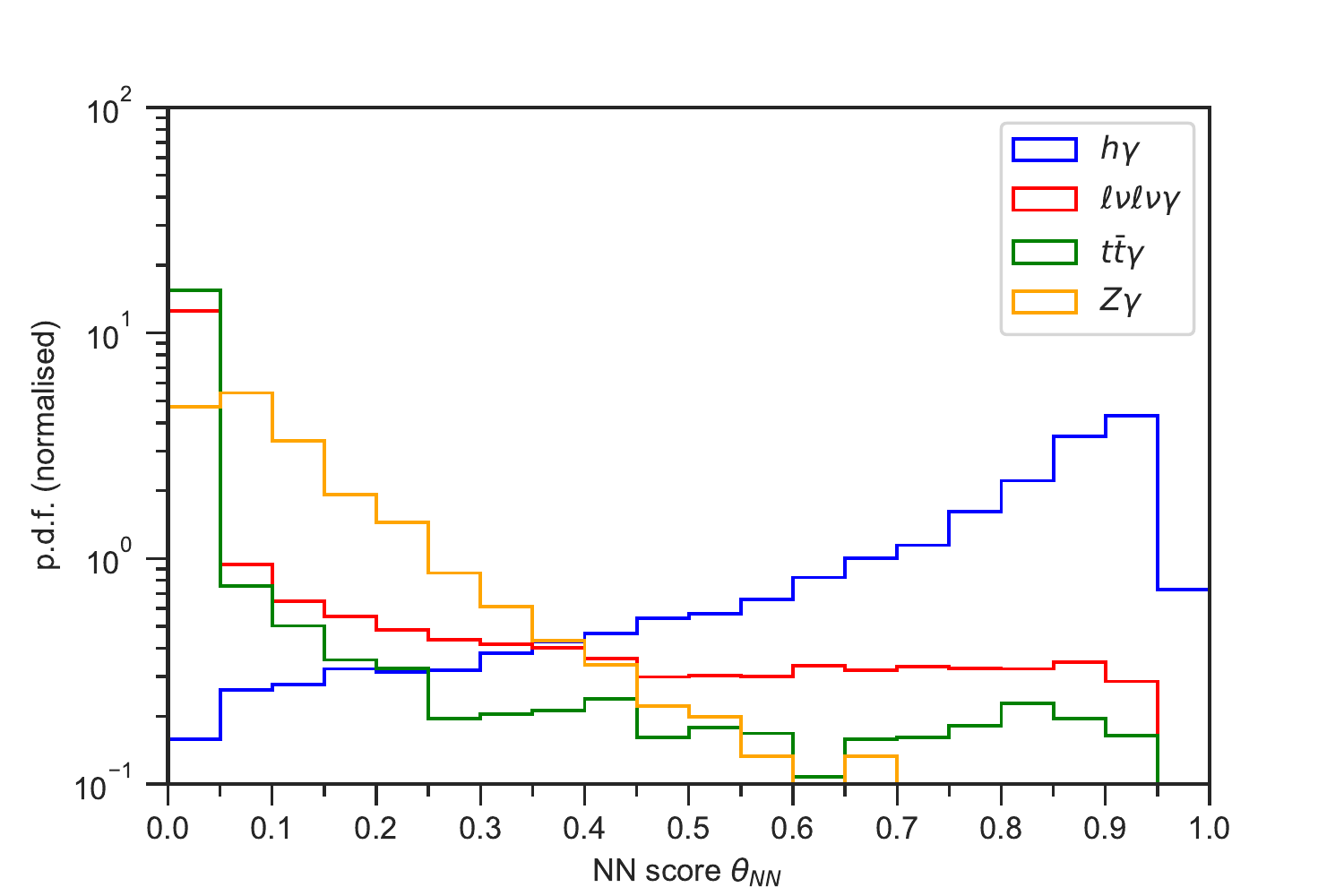}
\vspace{-4mm}
\caption{Multivariate NN score variable $\theta_\mathrm{NN}$ for the $h \gamma$ signal (blue) and dominant SM backgrounds $\ell^+\nu\ell^-\bar{\nu}\gamma$ (red), $t \bar{t} \gamma$ (green), and $Z (\to \tau^+ \tau^-) \gamma$ (yellow) in the charm-quark Yukawa sensitivity study.}
\label{fig:NN}
\par\end{centering}
\vspace{-2mm}
\end{figure}

The signal discrimination power achieved by our multivariate analysis is very high, with an area under the ``receiver operating characteristic'' (ROC) curve of 0.941 and 0.938 respectively for charm-quark and up-quark Yukawa sensitivity studies. The multivariate NN score variable $\theta_\mathrm{NN}$ (which may be regarded as a highly non-linear function of the kinematic variables in~\eqref{NN_variables}) for the signal and dominant SM backgrounds in the charm-quark Yukawa study is shown in Fig.~\ref{fig:NN}. In this case, a cut in the NN score variable $\theta_\mathrm{NN} > 0.78$ yields a signal efficiency $\sim 0.57$ together with SM background efficiencies $0.057$, $0.034$ and $0.003$ respectively for 
$\ell^+\nu\ell^-\bar{\nu}\gamma$, $t \bar{t} \gamma$ and $Z (\to \tau^+ \tau^-) \gamma$.
For the up-quark Yukawa study, the optimal cut is also found to be $\theta_\mathrm{NN} > 0.78$, yielding a signal efficiency $\sim 0.56$ and respective SM background efficiencies $0.056$, $0.031$ and $0.003$. 

\vspace{1mm}

In addition to the dominant SM backgrounds, we also consider the VBF and AP $h + \gamma$ production processes as potential, yet minor backgrounds for our charm and up-quark Yukawa sensitivity analysis, as discussed in section II. The extra high-$p_T$ activity vetoes imposed in our initial event selection suppress these processes down to a $h (\to \ell^+\nu\ell^-\bar{\nu})  \gamma$ cross section (assuming SM branching fractions for the Higgs boson) of $32.6$ ab for VBF, $2.24$ ab for $h\gamma W$ (with $W \to jj$ or $W \to \ell\nu$) and $1.84$ ab for $h\gamma Z$ (with $Z \to jj$ or $Z \to \nu \bar{\nu}$), with other backgrounds like $h \gamma j$ and $tth \gamma$ negligible after the event selection. Due to such small cross sections, these backgrounds are not 
included in the NN training. The NN selection efficiencies for them are the following: in the charm-quark Yukawa study, the cut $\theta_\mathrm{NN} > 0.78$ yields the efficiencies $0.42$, $0.25$ and $0.27$ for the VBF, $h\gamma W$ and $h\gamma Z$ backgrounds, respectively; for the up-quark Yukawa case, the cut $\theta_\mathrm{NN} > 0.78$ yields the corresponding efficiencies $0.42$, $0.26$ and $0.28$. Altogether, these backgrounds do not appreciably reduce the sensitivity to $\kappa_c$ and $\kappa_u$ from our multivariate analysis, which is driven by the NN ability to reject the main irreducible SM background, $p p \to \ell^+\nu\ell^-\bar{\nu}\gamma$.


%
%



\vspace{2mm}

\noindent \textbf{IV. Constraints on {\large $\kappa$}$_c$ \& {\large $\kappa$}$_u$.}~For SM branching fractions of the Higgs boson, the sensitivity to $\kappa_c$ and $\kappa_u$ at the HL-LHC from the NN analysis of the previous section is $|\kappa_c| < 11.8$ and $|\kappa_u| < 1930$ at 95\% C.L. 
(improving on the cut-and-count analysis from section III, as expected). This assumes that the statistical uncertainty of the SM background will largely dominate over its systematic uncertainty at the HL-LHC, which is justified in the present scenario, particularly since the main backgrounds are electroweak processes. The above projected bounds also assume that only one Yukawa coupling of the Higgs boson departs from its SM value.   

Enhancing $y_c$ or $y_u$ by an amount that makes them comparable to the SM bottom quark Yukawa coupling would modify significantly the total width of the Higgs boson and therefore its branching fractions.~Nevertheless, it has long been realized that    
light quark Yukawa couplings remain essentially unconstrained by global fits to Higgs production and decay rates at the LHC~\cite{Zeppenfeld:2000td,Duhrssen:2004cv,Belanger:2013xza} (see also~\cite{Coyle:2019hvs}), unless further assumptions are made. 
The effect of an enhanced Higgs Yukawa coupling $y_q$ to a light quark $q=u,d,c,s$ 
on the Higgs branching fractions may be compensated by a related increase of the Higgs couplings to gauge bosons and third-generation fermions, leading to a ``flat direction'' in the fit along which the Higgs signal strengths remain unchanged.
From the present good agreement between SM predictions and LHC Higgs measurements~\cite{CMS:2020gsy,Aad:2019mbh,ATLAS:2020wny}, this flat direction may be approximately described by a single generic $\kappa_h$ enhancement factor for all Higgs couplings other than the light quark Yukawa $y_q$ of interest~\cite{Coyle:2019hvs}
\begin{equation}
\kappa_h^2 \simeq \frac{1-Br_{q\bar{q}}^{\rm SM}}{2} +
\frac{\sqrt{(1-Br_{q\bar{q}}^{\rm SM})^2 + 4\, Br_{q\bar{q}}^{\rm SM} \,\kappa_q^2}}{2}\,,
    \label{eq:scaling}
\end{equation}
with $Br_{q\bar{q}}^{\rm SM}$ the branching fraction for $h \to q\bar{q}$ in the SM.
While the combination of Higgs signal strengths with other measurements, e.g. with electroweak precision observables or an indirect measurement of the Higgs total width (model dependent, 
see~\cite{Englert:2014aca}) can help lifting the flat direction~\eqref{eq:scaling}, this discussion highlights the importance of complementary probes of Higgs couplings to light quarks. 

Considering $\kappa_c$ and $\kappa_u$ along the flat direction defined by~\eq{eq:scaling} weakens our analysis' sensitivity w.r.t.~the assumption of SM branching fractions, since $\kappa_q > \kappa_h$ for $q = c, u$, and the effect of this becomes particularly important once $y_q/y^{\rm SM}_b \gtrsim 1$. The projected 95\% C.L. sensitivities to $\kappa_c$ and $\kappa_u$ along the flat direction are $|\kappa_c| < 26.3$ and $|\kappa_u| < 2300$. 

The projected bounds on $\kappa_c$ which we obtain are complementary to other existing probes in the literature. Yet, they may not be competitive with the most sensitive proposed direct probes of the charm Yukawa coupling~\cite{Brivio:2015fxa,Bishara:2016jga}, which yield a current $95\%$ C.L. experimental limit on $\kappa_c$ (assuming SM Higgs branching fractions) of $\kappa_c \lesssim 13$~\cite{ATLAS:2020wny}. In contrast, the achievable $h\, \gamma$ sensitivity to $\kappa_u$ does lie in the same ballpark of other currently proposed probes. 


%

\vspace{2mm}

\noindent \textbf{V. Conclusions.}~In this Letter we have studied $h \, \gamma$ production at the HL-LHC. While interesting in its own right, as this process remains yet to be observed at the LHC, we demonstrate its role as a sensitive probe of the Higgs boson couplings to the light quarks of the first two generations of matter, still largely unconstrained by present measurements.~The associated production with a photon enhances the contribution of the up-type quarks with respect to their down-type counterparts, yielding a way to disentangle Yukawa coupling enhancements from both quark types. This makes $h + \gamma$ highly complementary to other existing light quark Yukawa probes.~Concentrating on the $h \to \ell^+ \nu \ell^- \bar{\nu}$ decay channel of the Higgs boson, we have performed a multivariate neural network analysis to fully exploit the rich kinematics of this final state, and derived HL-LHC projected sensitivities to the Higgs Yukawa couplings to charm and up quarks. Particularly in the latter case, $h + \gamma$ may help to gain further insight on Higgs flavor at the LHC. 




\begin{acknowledgments}

\begin{center}
\textbf{Acknowledgements} 
\end{center}

\vspace{-2mm}

Feynman diagrams were drawn using {\sc TikZ-Feynman}~\cite{Ellis:2016jkw}.~J.A.A.S. acknowledges partial financial support by the Spanish ``Agencia Estatal de Investigaci\'on'' (AEI) through the project PID2019-110058GB-C21. The work of J.M.C. was supported by the Spanish MICIU and the EU Fondo Social Europeo (FSE) through the grant PRE2018-083563.
The work of J.M.N. was supported by the Ram\'on y Cajal Fellowship contract RYC-2017-22986, and by grant PGC2018-096646-A-I00 from the Spanish Proyectos de I+D de Generaci\'on de Conocimiento.
J.M.N. also acknowledges support from the European Union's Horizon 2020 research and innovation programme under the Marie Sklodowska-Curie grant agreement 860881 (ITN HIDDeN), as well as from 
the AEI through the grant IFT Centro de Excelencia Severo Ochoa SEV-2016-0597.

\end{acknowledgments}

\bibliography{Higgs_Photon_V2}

\end{document}